\documentclass[aps,prb,longbibliography,reprint,amsmath,amssymb,superscriptaddress]
{revtex4-2}
\usepackage{graphicx}
\usepackage{dcolumn}
\usepackage{bm}
\usepackage{color} 
\usepackage{physics}
\usepackage{hyperref}
\usepackage{subfigure}
\hypersetup{
    colorlinks = true,
    linkcolor = blue,
    citecolor = blue,
    anchorcolor = blue,
    urlcolor = blue
    }

\def\lesssim{\ \raise.3ex\hbox{$<$}\kern-0.8em\lower.7ex\hbox{$\sim$}\ }
\def\gesim{\ \raise.3ex\hbox{$>$}\kern-0.8em\lower.7ex\hbox{$\sim$}\ }
\def\up{\uparrow}

\newcommand \beq{\begin{eqnarray}}
\newcommand \eeq{\end{eqnarray}}

\usepackage{amsmath,amssymb}
\usepackage{fixmath}
\usepackage[normalem]{ulem}

\usepackage{comment}

\begin{document}

\preprint{RIKEN-iTHEMS-Report-26}

\title{Josephson effects in an interaction-asymmetric junction across the BCS-BEC crossover}

\author{Tingyu Zhang}
\email{zhangtv@hku.hk}
\affiliation{Department of Physics, State Key Laboratory of Optical Quantum Materials, and Hong Kong Institute of Quantum Science and Technology, University of Hong Kong, Hong Kong, China}
\author{Hiroyuki Tajima}
\affiliation{Department of Physics, Graduate School of Science, The University of Tokyo, Tokyo 113-0033, Japan}
\affiliation{RIKEN Nishina Center, Wako 351-0198, Japan}
\affiliation{Quark Nuclear Science Institute, The University of Tokyo, Tokyo 113-0033, Japan}

\begin{abstract}

We theoretically study the Josephson effect in ultracold Fermi gases, where the two sides of the Josephson junction are independently tuned to different regions of the Bardeen–Cooper–Schrieffer (BCS)–Bose–Einstein condensation (BEC) crossover. Using the nonequilibrium Green's function approach combined with the tunnel Hamiltonian formalism and a mean-field approach, we evaluate the DC and AC Josephson currents throughout the entire crossover region.
We calculate the DC Josephson current as a function of interaction strength by tuning both sides of the junction synchronously from the BCS to the BEC regimes, and give the asymptotic expression of the current in the deep BCS and BEC limits. We also study the AC Josephson junction through the interaction-asymmetric junction by fixing the interaction in one reservoir and tuning that of the other one. A peak of the tunneling current is found when one side is fixed in the BCS limit and the other side is tuned into the BEC regime, which corresponds to the interaction-biased Riedel peak. Our results indicate the competition between contributions of increasing pair spectral weight and decreasing chemical potential to Josephson tunneling throughout the BCS-BEC crossover, and demonstrate the realization of the Riedel peak in strong-coupling quantum gases.

\end{abstract}

\maketitle

\section{Introduction}

Josephson effect~\cite{JOSEPHSON1962251}, a fundamental quantum phenomenon arising in pairing systems, has provided one of the most direct macroscopic manifestations of quantum phase coherence, leading to profound implications for both fundamental physics and applied superconducting technology. Unlike normal electronic conduction, the Josephson current results from coherent tunneling of Cooper pairs and does not require an applied voltage. It can thus reveal the pairing natures of Cooper pairs both in solid state superconductors~\cite{RevModPhys.67.515,RevModPhys.72.969}, and neutral superfluids or condensates of ultracold atoms~\cite{RevModPhys.87.803,PhysRevLett.79.4950,PhysRevA.64.033610,PhysRevLett.95.010402,PhysRevLett.106.025302,PhysRevLett.118.230403}.
Apart from these systems, such an effect has also been discussed in nuclear matter~\cite{7bj7-b15c,PhysRevD.111.023044} and in nuclear reaction processes~\cite{PhysRevLett.119.042501,PhysRevC.105.L061602}, which provides inspiration for understanding the physical processes inside large nuclear systems such as neutron stars. 

With the advancement of experimental techniques, the Josephson effect has been observed in ultracold atomic systems~\cite{valtolina2015josephson,PhysRevLett.120.025302,science.aaz2463,science.aaz2342,science.ads9061}. In Fermi systems, an atomic Josephson junction has been  realized in ultracold $^6$Li atoms~\cite{science.aaz2342}, where a relative phase between different sides is imprinted by illuminating one of the reservoirs with an optical potential for a variable time that is much shorter than the Fermi time. Meanwhile, the BCS-BEC crossover can be tuned by Feshbach resonance between two hyperfine states of ultracold fermions through an external magnetic field~\cite{PhysRevLett.92.040403,PhysRevLett.92.120401,BCS-BEC}, allowing for studies of collective transport phenomena in different interacting regimes using one single system~\cite{pnas.1601812113,pnasnexus/pgad045,PhysRevB.108.155303,PhysRevApplied.21.L031001}. The transport properties of Cooper pairs can thus be studied through Josephson effect from weakly bound fermion pairs to tightly bound molecular-like pairs~\cite{PhysRevA.100.063601,PhysRevA.102.013325}.

Regarding the theoretical side, microscopic calculations for Josephson current based on Bogoliubov de Gennes (BdG) equations~\cite{PhysRevLett.99.040401,SPUNTARELLI2010111,PhysRevB.102.144517,PhysRevB.111.134517} have provided detailed descriptions of the transport in the crossover region.
While the BdG approach is useful to consider the microscopic spatial structure of the junctions, it is numerically demanding even in the weakly-interacting BCS regime.
Meanwhile, the nonequilibrium Green's function approach, developed by Schwinger and Keldysh~\cite{schwinger1961brownian,Keldysh},
and systematized by Kadanoff–Baym equations~\cite{PhysRev.124.287,PhysRev.127.1391},
has been applied to the study of tunneling transport in condensed-matter systems~\cite{zagoskin1998quantum,mahan2013many}.
An important advantage of the nonequilibrium Green's function approach is that one can investigate many-body aspects of the tunneling transport, where the geometric properties of the tunneling junctions are summarized into the tunneling amplitude. In this way, one can concisely obtain universal expressions of the tunneling transport in the weak-coupling limit, such as Ambegaokar-Baratoff formula~\cite{PhysRevLett.10.486},
Andreev information mirror~\cite{PhysRevD.96.124011,PhysRevD.102.064028,zhang2023dominant},
and Wiedemann-Franz law in degenerate fermions~\cite{PhysRevA.111.033312}).
Another advantage is its extendability to include multi-particle tunneling~\cite{PhysRevA.106.033310,pnasnexus/pgad045}, which plays a crucial role in dense matter~\cite{PhysRevD.111.023044} and itinerant ferromagnetic states~\cite{PhysRevB.108.155303,PhysRevApplied.21.L031001},
and is difficult to incorporate in the BdG approach.
In this sense, the formulation for quantum transport based on the nonequilibrium Green's function approach is
complementary to the BdG approach,
and thus important for understanding the Josephson current throughout the BCS-BEC crossover. Indeed, such techniques combined with the tunneling Hamiltonian formulation has a long history in the microscopic theory of superconducting tunnel junctions, including studies of AC Josephson currents and related nonlinear effects~\cite{osti,PhysRev.147.255,A.Barone}.

Furthermore, motivated by the recent experimental progress in ultracold atom physics,
it is interesting to consider the Josephson junction with an interaction asymmetry.
The spatial control of the interaction strength~\cite{PhysRevLett.122.040405}
enables us to study a tunneling junction in which the two sides are tuned to different regimes of the BCS-BEC crossover.
The asymmetric junction may be realized in condensed-matter systems by connecting weakly-coupled and strongly-coupled superconductors.
Along this direction, the proximity effect and the Kibble-Zurek mechanism in the normal-superfluid spatial junction have been discussed theoretically~\cite{PhysRevA.107.063314}.

On the other hand, the bias of the interaction strength between the two reservoirs should be distinguished from ordinary thermodynamic biases such as chemical potential and temperature differences. A thermodynamic bias is generally associated with a field conjugate to a conserved quantity, for example $\mu$ to particle number and $T$ to energy transport~\cite{PhysRev.135.A1505}. In contrast, the interaction coupling does not play such a role and cannot be regarded as a thermodynamic bias in the strict sense.
Although the expectation value of the corresponding interaction operator is closely related to Tan's contact~\cite{tan2008energetics,tan2008large,tan2008generalized}, as the contact is regarded as a conjugate thermodynamic variable of the inverse scattering length $a^{-1}$~\cite{PhysRevA.86.013626}, this conjugacy is different from that between $\mu$ and the particle number or that between \textcolor{blue}{$1/T$} and the energy.
While the latter quantities are conserved and obey local continuity equations, the interaction strength is an externally controlled parameter of the Hamiltonian, and the contact is a generalized force measuring short-range pair correlations rather than a conserved density. 
Accordingly, there is no conserved contact current that would mediate equilibration between reservoirs.
Nevertheless, the interaction-asymmetric junction provides an experimentally tunable nonequilibrium transport that reshapes the quasiparticle spectrum and modifies the Josephson tunneling dynamics.
In this sense, it serves as a useful control knob for transport, even though its status differs from that of conventional thermodynamic driving forces. In the deep BEC regime, where fermion pairs behave as tightly bound bosonic dimers, this interaction-related degree of freedom may become more closely connected to an approximately conserved dimer density, further motivating the investigation of Josephson transport under the interaction asymmetry.

In this work, we analyze the Josephson current
through the interaction-asymmetric junction
in an ultracold Fermi gas throughout the entire BCS-BEC crossover based on the Schwinger-Keldysh approach, combined with a tunneling Hamiltonian formulation. Similar studies on the DC Josephson current has been conducted in a symmetric Josephson junction~\cite{PhysRevA.100.063601} where two reservoirs are tuned synchronously from the BCS to BEC limits, and a maximum critical current is found near the unitary limit as a result of the competition between the increasing condensate fraction and a decrease of the chemical potential. 
The location of the extreme point should be different in an asymmetric junction, where interparticle scattering lengths in two reservoirs are tuned independently, inducing the AC Josephson current. 
We investigate the AC Josephson current under the interaction asymmetry between the two sides of the tunneling junction, where we fix one reservoir in the BCS limit while tuning the other one from BCS to BEC limit. 
Our results provide further insight into Josephson tunneling in the presence of the competition between the enhanced spectral weight and the increasing mismatch of the two Fermi surfaces. 

The outline of this paper is as follows: In Sec.~\ref{Sec2}, we present the theoretical model for the atomic Josephson junction and derive the tunneling current from the tunneling Hamiltonian. In Sec.~\ref{Sec3}, we investigate the DC Josephson current without chemical potential bias in a symmetric junction with varying interaction strength. In Sec.~\ref{Sec4}, we calculate the AC Josephson current in an asymmetric junction involving the nonzero chemical-potential bias by fixing one side in the BCS limit and tuning the other side through the BCS-BEC crossover. Finally we summarize this paper and give perspectives in Sec.~\ref{Sec5}.

\section{Tunneling model}\label{Sec2}
Throughout the paper, we take $\hbar = k_B =1$ and the volumes for both reservoirs to be unity.
We consider two-component Fermi gases trapped in a two-terminal model, where the scattering length in each side can be tuned independently through Feshbach resonances. The effective Hamiltonian is given by
\begin{align}
    \hat{H}=\hat{H}_{\rm L}+\hat{H}_{\rm R}+\hat{H}_{\rm T},
\end{align}
including the reservoir Hamiltonian $\hat{H}_{i={\rm L},{\rm R}}$ and tunneling Hamiltonian $\hat{H}_{\rm T}$. Notice that although other terms associated with particle reflection and induced interface interaction also arise due to the potential barrier and the two-body interaction~\cite{PhysRevA.106.033310}, we omit them since they are irrelevant to our consideration.
The reservoir Hamiltonian is given by
\begin{align}
    \hat{H}_{\rm L}=&\sum_{\bm{k},\sigma}\xi_{\bm{k},{\rm L}}c^\dagger_{\bm{k},\sigma,\rm{L}}c_{\bm{k},\sigma,\rm{L}}\cr
    +&V_{\rm L}\sum_{\bm{k},\bm{k}',\bm{q}}c^\dagger_{\bm{k}+\bm{q},\up,\rm{L}}c^\dagger_{-\bm{k},\downarrow,\rm{L}}c_{-\bm{k}'\downarrow,\rm{L}}c_{\bm{k}'+\bm{q},\up,\rm{L}},
\end{align}
\begin{align}
    \hat{H}_{\rm R}=&\sum_{\bm{k},\sigma}\xi_{\bm{k},{\rm R}}c^\dagger_{\bm{k},\sigma,\rm{R}}c_{\bm{k},\sigma,\rm{R}}\cr
    +&V_{\rm R}\sum_{\bm{k},\bm{k}',\bm{q}}c^\dagger_{\bm{k}+\bm{q},\up,\rm{R}}c^\dagger_{-\bm{k},\downarrow,\rm{R}}c_{-\bm{k}'\downarrow,\rm{R}}c_{\bm{k}'+\bm{q},\up,\rm{R}},
\end{align}
where $\xi_{\bm{k},i}=p^2/(2m)-\mu_{i}$, $c_{\bm{k},\sigma,i}$ denotes the particle annihilation operator, and $V_i$ is the s-wave interaction strength associated with the scattering length $a_i$ in reservoir $i$. The tunneling Hamiltonian is given by
\begin{align}\label{eq:HT}
    \hat{H}_{\rm T}=\sum_{\bm{k},\bm{k}',\sigma}\mathcal{T}_{\bm{k},\bm{k}',\sigma}c^\dagger_{\bm{k},\sigma,\rm{R}}c_{\bm{k}',\sigma,\rm{L}}+\rm{h.c.},
\end{align}
where $\mathcal{T}_{\bm{k},\bm{k}',\sigma}$ is the tunneling coupling strength. Usually we consider the barrier potential as a delta potential $V(z)=V_0\delta(z/\lambda)$ yielding a constant $V(\bm{k})=V_0$ in the momentum space or a rectangular potential barrier reading $V(z)=0$ for $z < 0$ and $z > \lambda$, and $V(z)=V_0$ for $0 \leq z \leq \lambda$. With the assumption of momentum conservation during the tunneling process ($\bm{k}-\bm{k}'\rightarrow0$), the tunneling strengths can be represented as $\mathcal{T}_{\bm{k},\bm{k}',\sigma}\equiv\mathcal{T}_{\bm{k},\sigma}\delta_{\bm{k}\bm{k}'}=[C_{\bm{k},\sigma}\epsilon_k+V_0\mathcal{B}_{k_z,\sigma}]\delta_{\bm{k}\bm{k}'}$~\cite{PhysRevA.106.033310}, where $C_{\bm{k},\sigma}$ denotes the transmission coefficient of one particle tunneling through the barrier and $\mathcal{B}_{k_z,\sigma}$ is the overlap integral of wave functions inside the potential barrier. Its momentum-dependence and barrier dependence can thus be analyzed at the microscopic level, which is studied in Ref.~\cite{PhysRevApplied.21.L031001}. For an ideal translationally invariant barrier, momentum parallel to the interface is conserved, and in a simplified momentum conserved model we write the tunneling coupling strength with a factor $\delta_{\bm{k}\bm{k}'}$.

Based on the Heisenberg equation, we introduce the tunneling current operator 
\begin{align}
    \hat{I}=-\frac{d}{dt}\hat{N}_{\rm L}=i\big[\hat{N}_{\rm L},\hat{H}_{\rm T}\big],
\end{align}
where $\hat{N}_i=\sum_{\bm{k},\sigma}c^\dagger_{\bm{k},\sigma,i} 
c_{\bm{k},\sigma,i}$ is the number density operator. Notice that $\hat{N}$ is commutative with $\hat{H}_{\rm L}$ and $\hat{H}_{\rm R}$. Thus its changing rate is given by the commutation of $\hat{N}_{\rm L}$ and $\hat{H}_{\rm T}$. Then the current operator can be written as 
\begin{align}
   \hat{I}=-i\sum_{\bm{k},\bm{k}',\sigma}\mathcal{T}_{\bm{k},\bm{k}',\sigma}c^\dagger_{\bm{k},\sigma,\rm{R}}c_{\bm{k}',\sigma,\rm{L}}+\rm{h.c.},
\end{align}
which represents the changing rate of particle number in the left reservoir due to the tunneling toward the right side. We are interested in its expectation value, which is given by
\begin{align}\label{Iexpectation}
    I(t,t')=&\sum_{n=0}^{\infty}\frac{(-i)^n}{n!}\int_Cdt_1\cdots\int_Cdt_n \cr
    &\times\langle {\rm T}_C \hat{I}(t,t')\hat{H}_{\rm T}(t_1)\cdots \hat{H}_{\rm T}(t_n)\rangle.
\end{align}
The time parameter $t$ and $t'$ respectively locate on the backward (from $t=+\infty$ to $t=-\infty$) and forward (from $t=-\infty$ to $t=+\infty$) branches of the Keldysh contour $C$, and $T_C$ is the contour ordering operator which places the operator with time argument locating further along the contour earlier in the order.
We assume the total system is in a non-equilibrium steady state, where current flows from one reservoir to another while each reservoir is large enough to hold local equilibrium state inside. Thus grand-canonical Hamiltonian should be taken into account as calculating correlation functions for particles in locally equilibrium state. Thus we take the transform for annihilation and creation operators:
\begin{align}
    &c^\dagger_{\bm{k},\sigma,i}(t)\rightarrow e^{i\mu_it}c^\dagger_{\bm{k},\sigma,i}(t),\cr 
    &c_{\bm{k},\sigma,i}(t)\rightarrow e^{-i\mu_it}c_{\bm{k},\sigma,i}(t).
\end{align}
Then Eq.~(\ref{Iexpectation}) becomes 
\begin{align}
    I(&t,t')=-i\sum_{\bm{k},\bm{k'},\sigma}\mathcal{T}_{\bm{k},\bm{k}',\sigma}\sum_{n=0}^{\infty}\frac{(-i)^n}{n!}\int_Cdt_1\cdots\int_Cdt_n\nonumber\\
    &\langle {\rm T}_C e^{i\mu_{\rm R}t}e^{-i\mu_{L}t'}c^\dagger_{\bm{k},\sigma,\rm{R}}(t)c_{\bm{k}',\sigma,\rm{L}}(t')\hat{H}_{\rm T}(t_1)\cdots \hat{H}_{\rm T}(t_n)\rangle.
\end{align}
Intercepting to the leading-order term, we have
\begin{align}\label{I1st}
    I(t,t')=&-\sum_{\bm{k},\bm{k}',\sigma}\mathcal{T}_{\bm{k},\bm{k}',\sigma}\int_C dt_1 \nonumber\\
    &\langle {\rm T}_C e^{i\mu_{\rm R}t}e^{-i\mu_{\rm L}t'}c^\dagger_{\bm{k},\sigma,\rm{R}}(t)c_{\bm{k}',\sigma,\rm{L}}(t')\hat{H}_{\rm T}(t_1)\rangle.
\end{align}
We introduce the Green's function within the Nambu representation
\begin{align}
    i\mathcal{G}_{\bm{k},\sigma,i}(t,t')=
    \begin{pmatrix}
    G_{\bm{k},\sigma,i}(t,t') & F_{\bm{k},i}(t,t') \\
    F^{\dagger}_{\bm{k},i}(t,t') & G^\dagger_{\bm{k},\sigma,i}(t,t')
\end{pmatrix}
\end{align}
The diagonal propagators are defined as 
\begin{align}
    &G_{\bm{k},\sigma,i}(t,t')=\big\langle {\rm T}_Cc_{\bm{k},\sigma,i}(t)c^\dagger_{\bm{k},\sigma,i}(t')\big\rangle,\nonumber\\
    &G^\dagger_{\bm{k},\sigma,i}(t,t')=\big\langle {\rm T}_Cc_{-\bm{k},\bar{\sigma},i}^\dagger(t)c_{-\bm{k},\bar{\sigma},i}(t')\big\rangle,
\end{align}
where $\bar{\sigma}$ denotes the opposite spin of $\sigma$.
The off-diagonal elements represent the abnormal propagators, arising from the non-conservation nature of particle number in pair condensate,
\begin{align}
    &F_{\bm{k},i}(t,t')=\big\langle {\rm T}_Cc_{\bm{k},\uparrow,i}(t)c_{\bm{k},\downarrow,i}(t')\big\rangle,\nonumber\\
    &F^\dagger_{\bm{k},i}(t,t')=\big\langle {\rm T}_Cc^\dagger_{\bm{k},\downarrow,i}(t)c^\dagger_{\bm{k},\uparrow,i}(t')\big\rangle.
\end{align}
Here ${\rm T}_C$ orders the operators in sequence along the Keldysh contour $C$---on the forward branch ${\rm T}_C={\rm T}$, and on the backward branch ${\rm T}_C=\bar{\rm T}$. Setting $t=t'$, Eq.~(\ref{I1st}) can be rewritten as $I(t)=I_{\rm q}(t)+I_{\rm J}(t)$, where
\begin{widetext}
\begin{align}~\label{Iq}
    I_q=-2\sum_{\bm{k},\bm{k}',\sigma}\operatorname{Re}\bigg\{\mathcal{T}_{\bm{k},\bm{k}',\sigma}^2\int_{-\infty}^{\infty} dt_1 e^{-i\Delta\mu(t-t_1)}\Big[G^{\rm ret.}_{\bm{k}',\sigma,{\rm L}}(t,t_1) G^{<}_{\bm{k},\sigma,{\rm R}}(t_1,t)
    +G^<_{\bm{k}',\sigma,{\rm L}}(t,t_1) G^{\rm adv.}_{\bm{k},\sigma,{\rm R}}(t_1,t)\Big]\bigg\},
\end{align}
\begin{align}\label{IJ}
    I_{\rm J}=2\sum_{\bm{k},\bm{k}',\sigma}\operatorname{Re}\bigg\{\mathcal{T}^2_{\bm{k},\bm{k}',\sigma}\int_{-\infty}^{\infty} dt_1 e^{-i[\Delta\mu(t+t_1)+\Delta\phi]}\Big[F^{{\rm ret.}}_{\bm{k}',{\rm L}}(t,t_1) F^{\dagger<}_{\bm{k},{\rm R}}(t_1,t)
    +F^{<}_{\bm{k}',{\rm L}}(t,t_1) F^{\dagger{\rm adv.}}_{\bm{k},{\rm R}}(t_1,t)\Big]\bigg\}.
\end{align}

\end{widetext}
Here $I_{q}$ represents the quasiparticle current while $I_{\rm J}$ represents the Josephson current. Notice that the integral along the Keldysh contour in Eq.~(\ref{I1st}) is replaced by the integral along the real time axis, according to Langreth rules. $G^{\rm ret.}$, $G^{\rm adv.}$ and $G^<$ are respectively retarded, advanced, and lesser Green's function. The chemical potential bias is defined as $\Delta\mu=\mu_{\rm L}-\mu_{\rm R}$. The gap parameter in each reservoir reads $\Delta_i=|\Delta_i|e^{-i\phi_i}$, and the phase different in Eq.~(\ref{IJ}) is defined by $\Delta\phi=\phi_{\rm L}-\phi_{\rm R}$.

\section{DC Josephson current}\label{Sec3}

Now we consider the interaction symmetric junction of which the two sides are tuned synchronously, where the DC Josephson current can arise without chemical potential barrier.
We rewrite the Josephson current Eq.~(\ref{IJ}) in the frequency representation:
\begin{widetext}
    \begin{align}\label{IJ2}
    I_{\rm J}=&4\sum_{\bm{k},\bm{k}',\sigma}\mathcal{T}^2_{\bm{k},\bm{k}',\sigma}\int_{-\infty}^{\infty} \frac{d\omega}{2\pi} \bigg\{\operatorname{Im}F_{\bm{k}',{\rm L}}(\omega) \operatorname{Im}F^{\dagger}_{\bm{k},{\rm R}}(\
    \omega-\Delta\mu)\big[f(\omega-\Delta\mu)-f(\omega)\big]\cos(2\Delta\mu t+\Delta\phi)\nonumber\\
    &-\Big[\operatorname{Re}F_{\bm{k}',{\rm L}}(\omega)\operatorname{Im}F^{\dagger}_{\bm{k},{\rm R}}(\omega-\Delta\mu)f(\omega-\Delta\mu)+\operatorname{Im}F_{\bm{k}',{\rm L}}(\omega)\operatorname{Re}F^{\dagger}_{\bm{k},{\rm R}}(\omega-\Delta\mu)f(\omega)\Big]\sin(2\Delta\mu t+\Delta\phi)\bigg\},
\end{align}
\end{widetext}
where we omit the superscript ${\rm ret.}$ for all retarded Green's functions. $f(\omega)=1/(e^{\omega/T}+1)$ is the Fermi distribution function.
The DC Josephson current appears when $\Delta\mu=0$ and $\Delta\phi\neq0$:
\begin{align}
    I_{\rm DC}=-4&\sum_{\bm{k},\bm{k}',\sigma}\mathcal{T}^2_{\bm{k},\bm{k}',\sigma}\int_{-\infty}^{\infty} \frac{d\omega}{2\pi}\Big[\operatorname{Re}F_{\bm{k}',{\rm L}}(\omega)\operatorname{Im}F^{\dagger}_{\bm{k},{\rm R}}(\omega)\nonumber\\
    &+\operatorname{Im}F_{\bm{k}',{\rm L}}(\omega)\operatorname{Re}F^{\dagger}_{\bm{k},{\rm R}}(\omega)\Big]f(\omega)\sin(\Delta\phi).
\end{align}
The imaginary part of the abnormal propagators is given by
\begin{align}
    &\operatorname{Im}F_{\bm{k},i}(\omega)=\operatorname{Im}F^{\dagger}_{\bm{k},i}(\omega)\cr
    &=\frac{\pi|\Delta_i|}{2E_{\bm{k},i}}\big[\delta(\omega-E_{\bm{k},i})-\delta(\omega+E_{\bm{k},i})\big]
\end{align}
where $E_{\bm{k},i}=\sqrt{\xi_{\bm{k},i}^2+|\Delta_i|^2}$ is the BCS quasiparticle excitation dispersion. first we assume the same condensate in both reservoirs in which case they have the same gap energy $|\Delta_{\rm L}|=|\Delta_{\rm R}|=|\Delta|$ and chemical potential $\mu_{\rm L}=\mu_{\rm R}=\mu$, and therefore the same dispersion $E_{\bm{k},{\rm L}}=E_{\bm{k},{\rm R}}=E_{\bm{k}}$. The DC Josephson current can be rewritten as 
\begin{widetext}
    \begin{align}
    I_{\rm DC}=&2\sum_{\bm{k},\bm{k}',\sigma}\frac{|\Delta|^2}{E_{\bm{k}}E_{\bm{k}'}}\mathcal{T}^2_{\bm{k},\bm{k}',\sigma}
    \bigg(\frac{1}{E_{\bm{k}}-E_{\bm{k}'}}-\frac{1}{E_{\bm{k}}+E_{\bm{k}'}}\bigg)\Big[2f(E_{\bm{k}})-1\Big]\sin{(\Delta\phi)}\nonumber\\
    =&2\sum_{\bm{k},\bm{k}',\sigma}\frac{|\Delta|^2[1-(f(E_{\bm{k}})+f(E_{\bm{k}'}))]}{E_{\bm{k}}E_{\bm{k}'}(E_{\bm{k}}+E_{\bm{k}'})}\mathcal{T}^2_{\bm{k},\bm{k}',\sigma}\sin{(\Delta\phi)}
    +2\sum_{\bm{k},\bm{k}',\sigma}\frac{|\Delta|^2}{E_{\bm{k}}E_{\bm{k}'}}\mathcal{T}^2_{\bm{k},\bm{k}',\sigma}
    \frac{f(E_{\bm{k}})-f(E_{\bm{k}'})}{E_{\bm{k}}-E_{\bm{k}'}}\sin{(\Delta\phi)}.
\end{align}

\end{widetext}

\begin{figure}[t]
    \includegraphics[width=8.5cm]{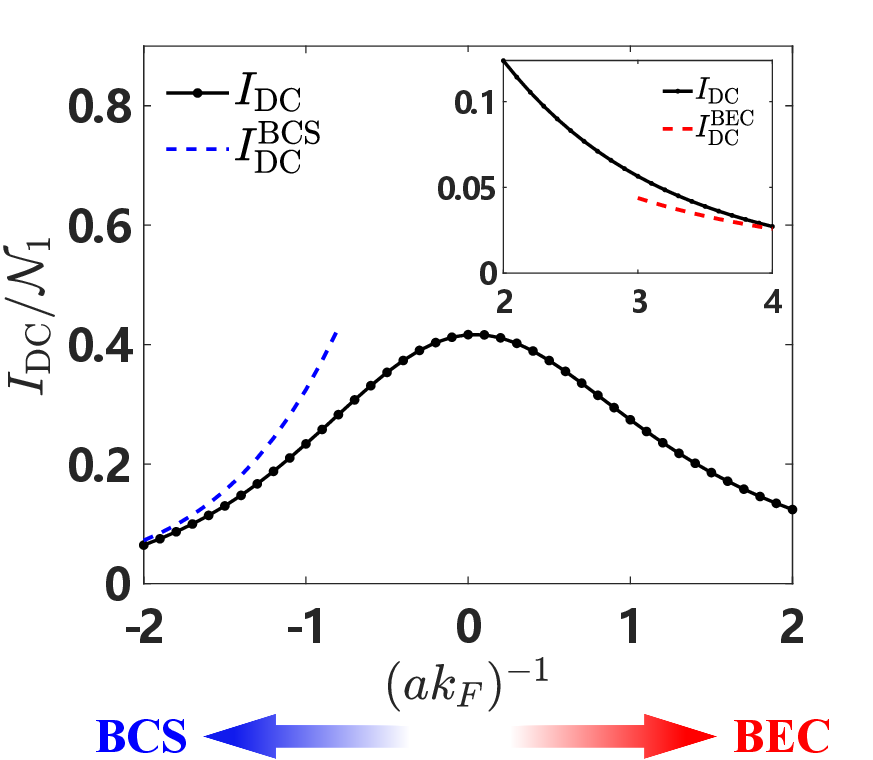}
    \caption{The DC Josephson current as a function of interaction strength (the black solid line) when the system changes from BCS to BEC regime. The tunneling is between two condensates with the same interaction strength, which is tuned from BCS to BEC limit. The momentum cutoff for the tunneling coupling strength is chosen to be $\Lambda=2k_{\rm F}$. $\mathcal{N}_1=9\mathcal{T}^2N^2\sin{(\Delta\phi)}/\epsilon_{\rm F}$ is the normalizing factor with $N=k^3_{\rm F}/(3\pi^2)$ denoting the number density. The blue dashed line shows the asymptotic behavior of the DC current in BCS limit. The inset figure shows the numerical result (black solid line) and the asymptotic analytical solution (red dashed line) of the DC current in deeper BEC limit, with $2<(ak_{\rm F})^{-1}<4$. 
    }
    \label{IDC-V}
\end{figure}

We calculate the DC Josephson current as we tune both reservoirs from BCS to BEC limit. We notice that the divergence arises from the integral over the high-momentum region. To avoid the divergence, we introduce a momentum cutoff into the tunneling coupling strength by rewriting $\mathcal{T}_{\bm{k},\bm{k}',\sigma}$ as
\begin{align}
\label{eq:20}
    \mathcal{T}_{\bm{k},\bm{k}',\sigma}=\frac{\mathcal{T}}{\sqrt{1+(k/\Lambda)^2}\sqrt{1+(k'/\Lambda)^2}},
\end{align}
where $\Lambda$ is a high momentum cutoff imposed through a momentum-dependent form factor~\cite{PhysRevA.64.033610}. Notice here we use the model for noncondensate-noncondensate tunneling amplitude since the tunneling Hamiltonian is written in terms of single-fermion operators as Eq.~(\ref{eq:HT}), which reflects the fact that the microscopic barrier acts on individual atoms. This should be distinguished from the condensate-condensate tunneling process in a weakly interacting Bose gas, where the bosonic field has a macroscopic single-particle condensate amplitude. Also the phenomenological single-fermion tunneling form factor with a smooth high-momentum cutoff is different from the case in Ref.~\cite{PhysRevApplied.21.L031001} where the momentum dependence of $\mathcal{T}$ is studied microscopically and a tunneling coupling constant $\mathcal{T_{k_{\rm F},k_{\rm F}}}$ is chosen as an assumption.
During the BCS-BEC crossover, the magnitude of the gap parameter $|\Delta|$ changes while the phase of the gap parameter remain unchanged, as the symmetry-broken state is unchanged. We plot the current-interaction characteristic at $T=0$ in Fig.~{\ref{IDC-V}}. The value of the chemical potential and the gap parameter is obtained by the BCS-Eagles-Legget theory, namely the mean-field theory for the BCS-BEC crossover, where the attractive interaction is decoupled in the pairing channel by introducing the order parameter $\Delta=-V\sum_{\mathbf{k}}\langle c_{-\mathbf{k},\downarrow}c_{\mathbf{k},\uparrow}\rangle$. For each interaction strength $(k_Fa)^{-1}$, values of $\mu$ and $|\Delta|$ are obtained self-consistently from the regularized gap equation and number equation at fixed density. These mean-field results can be found in e.g., Ref.~\cite{OHASHI2020103739}. The momentum cutoff is chosen to be $\Lambda=2k_{\rm F}$ as a representative value for which the tunneling amplitude is almost constant for low-energy fermions around the Fermi surface, while high-momentum components are smoothly suppressed. Although the interaction effects introduce additional length scales, especially on the BEC side where the pair size and molecular character become important, in our formulation these many-body effects enter through the Green’s functions, gap parameter, and chemical potential, whereas the detailed modification of the bare tunneling form factor by the composite molecular structure is not explicitly resolved. For this reason, the precise value of the current, especially in the deep BEC regime, depends quantitatively on $\Lambda$, while the qualitative crossover behavior discussed here is robust against moderate changes of the cutoff. In the interaction symmetric junction, the DC current exhibits a maximum near the unitary limit as a result of the competition between the increasing gap energy and the decreasing chemical potential, which is qualitatively consistent with the result in~\cite{PhysRevA.100.063601}. Moreover, this peaked structure has been experimentally observed in an ultracold $^6$Li atomic gas with a tunable Josephson junction~\cite{science.aaz2463}.
We note that the qualitative behavior of $I_{\rm DC}$ do not depend on the value of $\Lambda$ and discuss the quantitative effects below.

To understand the peaked structure of $I_{\rm DC}$,
hereafter we discuss the asymptotic expression of $I_{\rm DC}$ in the BCS and BEC limits.
First,
in the BCS limit, $\mu$ is positive and $|\Delta|\ll\mu\approx\epsilon_{\rm F}$. Near the Fermi surface, $E_{\bm{k}}\sim|\Delta|$ and the denominator is small, while far from the Fermi surface $E_{\bm{k}}\approx|\xi_{\bm{k}}|$ and the integrand is suppressed like $|\Delta|^2/|\xi_{\bm{k}}|^3$. Therefore, the dominant contribution comes from $k\simeq k'\simeq k_{\rm F}$. To leading order we may set 
\begin{align}\label{TF}
    \mathcal{T}_{\bm{k},\bm{k}',\sigma}\simeq\mathcal{T}_{\rm F}\equiv \frac{\mathcal{T}}{1+(k_{\rm F}/\Lambda)^2},
\end{align}
and replace the momentum sums by energy integrals with the density of states at the Fermi surface: $\sum_{\bm{k}}\rightarrow N(0)\int d\xi$,
where $N(0)=mk_{\rm F}/2\pi^2$ is the density of states per spin component at the Fermi level. Then
\begin{align}\label{IBCS}
    I_{\rm DC}^{\rm BCS}\simeq 4\mathcal{T}_{\rm F}^2\sin(\Delta\phi)\int_{-\infty}^{\infty} d\xi \int_{-\infty}^{\infty} d\xi'\frac{N(0)^2|\Delta|^2}{E E'(E+E')},
\end{align}
where $E=\sqrt{\xi^2+|\Delta|^2}$ and $E'=\sqrt{\xi'^2+|\Delta|^2}$. Finally we obtain the asymptotic formula for the DC Josephson current in BCS limit (see Appendix~\ref{appendixA}):
\begin{align}
    I_{\rm DC}^{\rm BCS}\simeq4\pi^2 N(0)^2\mathcal{T}_{\rm F}^2|\Delta|\sin(\Delta\phi),
\end{align}
which is consistent with the Ambegaokar–Baratoff formula~\cite{PhysRevLett.10.486}. According to Eq.~(\ref{TF}), the BCS asymptotic expression is cutoff-dependent through the effective Fermi surface tunneling amplitude $\mathcal{T}_{\rm F}$. Such dependence vanishes when $k_F / \Lambda\rightarrow 0$ and $\mathcal{T}_{\rm F}\rightarrow\mathcal{T}$.
We show this asymptotic behavior in the BCS limit as the blue dashed line in Fig.~\ref{IDC-V}.
With increasing the interaction strength, $I_{\rm DC}^{\rm BCS}$ increases due to the enhancement of $|\Delta|$.

In the BEC regime, the asymptotic behavior of the chemical potential is given by $\mu\simeq -\epsilon_{\rm b}/2$~\cite{OHASHI2020103739}, where $\epsilon_{\rm b}=1/(ma^2)$ is the two-body binding energy.
In deep BEC limit $|\epsilon_{\rm b}|$ is large and thus one has a large negative $\mu$ with $|\mu|\gg|\Delta|$. We can then approximate the dispersions as
$E_{\bm{k}}\simeq\xi_{\bm{k}}+\frac{|\Delta|^2}{2\xi_{\bm{k}}}$
and $f(E_{\bm{k}})\rightarrow 0$.
The leading term of the current can be written as (see Appendix~\ref{appendixA})
\begin{align}
\label{eq:24}
    I_{\rm DC}^{\rm BEC}\simeq&\frac{4m^3|\Delta|^2\mathcal{T}^2\Lambda^4\sin(\Delta\phi)}{\pi^4}\nonumber\\
    &\times\frac{\Lambda^4\mathcal{J}_{\Lambda\Lambda}-2\Lambda^2\kappa^2\mathcal{J}_{\Lambda,\kappa}+\kappa^4\mathcal{J}_{\kappa,\kappa}}{(\Lambda^2-\kappa^2)^2},
\end{align}
where $\kappa=\sqrt{2m|\mu|}$.
In addition, we introduce 
\begin{align}
    \mathcal{J}_{\kappa\kappa}=\frac{\pi(\pi-2)}{8\kappa^4},
\end{align}
\begin{align}
    \mathcal{J}_{\Lambda\kappa}=\frac{\pi^2(\Lambda-2\kappa)}{8\Lambda\kappa^2(\Lambda^2-\kappa^2)}+\frac{\pi\arctan{\sqrt{\frac{\sqrt{2}\kappa-\Lambda}{\sqrt{2}\kappa+\Lambda}}}}{\Lambda(\Lambda^2-\kappa^2)\sqrt{2\kappa^2-\Lambda^2}},
\end{align}
\begin{align}
    \mathcal{J}_{\Lambda\Lambda}=\frac{\pi(8\Lambda\theta-2\pi\Lambda+\pi\sqrt{2\kappa^2-\Lambda^2})}{8\Lambda^2\sqrt{2\kappa^2-\Lambda^2}(\kappa^2-\Lambda^2)},
\end{align}
where $\theta=\arctan\big(\frac{\Lambda}{\sqrt{2}\kappa+\sqrt{2\kappa^2-\Lambda^2}}\big)$. We plot the asymptotic behavior of DC current in deep BEC limit ($(k_{\rm F}a)^{-1}>3$) as the red dashed line in the inset figure of Fig.~\ref{IDC-V}.
In contrast to $I_{\rm DC}^{\rm BCS}$,
$I_{\rm DC}^{\rm BEC}$ has a moderate $\Lambda$ dependence and decreases with increasing the interaction strength due to the change of $|\mu|$ rather than $|\Delta|$.
We note that Eq.~\eqref{eq:24} is valid only in the deep BEC limit as $I_{\rm DC}^{\rm BEC}$ diverges at $\Lambda=\kappa\simeq 1/a$. In this sense, the Josephson current in the BEC regime quantitatively depends on the junction geometry incorporated to $\mathcal{T}_{\bm{k},\bm{k}'}$ in contrast to that in the BCS regime. Nevertheless, our result can reproduce the qualitative peaked behavior of $I_{\rm DC}$ observed in the experiment~\cite{science.aaz2463}, even with the separable tunneling amplitude given by Eq.~\eqref{eq:20}.

In this way, combining these asymptotic expressions in the BCS and BEC limits,
we can qualitatively understand the peaked behavior of $I_{\rm DC}$ where the interplay of $|\Delta|$ and $|\mu|$ plays a crucial role. In the BCS limit, increasing the interaction strength enhances the magnitude of gap parameter and increases the Josephson current. In contrast, in the deep BEC limit with negative chemical potential, increasing $(k_Fa)^{-1}$ increases $|\mu|$ and suppresses the tunneling contribution, and thus $I_{\rm DC}^{\rm BEC}$ decreases. These opposite monotonic behaviors imply that a maximum should appear in the crossover region.

\begin{figure*}[t]
    \centering
    \includegraphics[width=0.7\hsize]{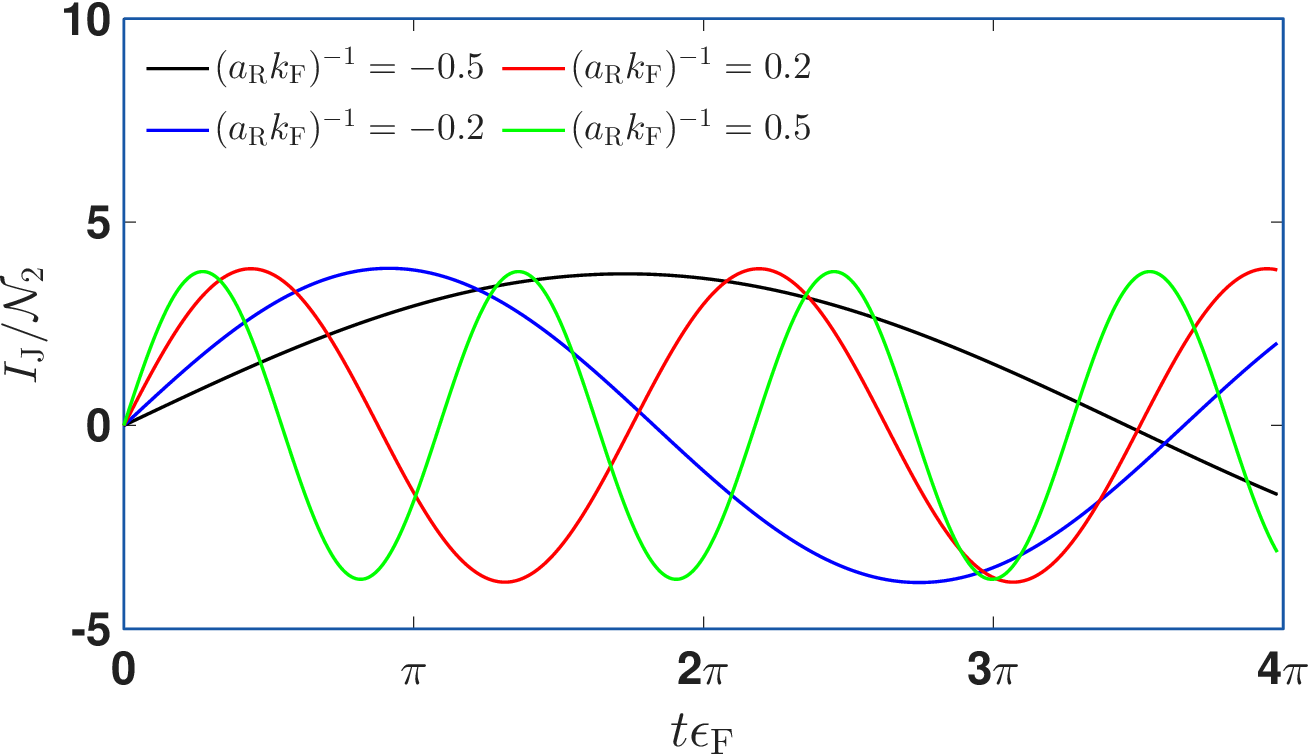}
    \caption{The AC Josephson current with different interaction strengths in the right reservoir, while the interaction strength in the left side is fixed as $(k_{\rm F}a_{\rm L})^{-1}=-1$. The top left corner shows the values of interaction strength in the right side. $\mathcal{N}_2=9\mathcal{T}^2|\Delta_{\rm L}|N^2/(4\epsilon_{\rm F}^2)$ is the normalizing factor. The initial phase bias is taken to be $\Delta\phi=0$ to minimize the Josephson energy. The periods of current are proportional to the reciprocal of chemical potential bias $1/\Delta\mu$. The momentum cutoff is chosen to be $\Lambda=2k_{\rm F}$.
    The time $t$ is normalized by the Fermi energy $\epsilon_{\rm F}$ in the left reservoir.
    }
    \label{IJ-t}
\end{figure*}

\section{Josephson current between different condensate regime}\label{Sec4}

We then consider the Josephson effect between two condensates where each side can be independently tuned in the BCS-BEC crossover, where the chemical potential bias is generally nonzero. In this section, we assume that the two reservoirs have the same density $N_{\rm L}=N_{\rm R}=N$, so that a common Fermi wave vector $k_{\rm F}=(3\pi^2N)^{1/3}$ and Fermi energy $\epsilon_{\rm F}=k_{\rm F}^2/(2m)$ can be used. The interaction strengths in the left and right reservoir are respectively characterized by the scattering lengths $a_{\rm L}$ and $a_{\rm R}$.
According to Eq.~(\ref{IJ2}) the DC Josephson current has a period proportional to $1/\Delta\mu$. First we fix the left side in the BCS limit with $1/(a_{\rm L}k_{\rm F})=-1$. The AC current at $T=0$ is shown in Fig.~\ref{IJ-t}. As the interaction of right side increases, both the period and magnitude of current changes due to the change of $\mu_{\rm R}$ and $|\Delta_{\rm R}|$. The former decreases due to the increasing of $\Delta\mu$. 
\begin{figure}[t]
    \centering
    \includegraphics[width=8.6cm]{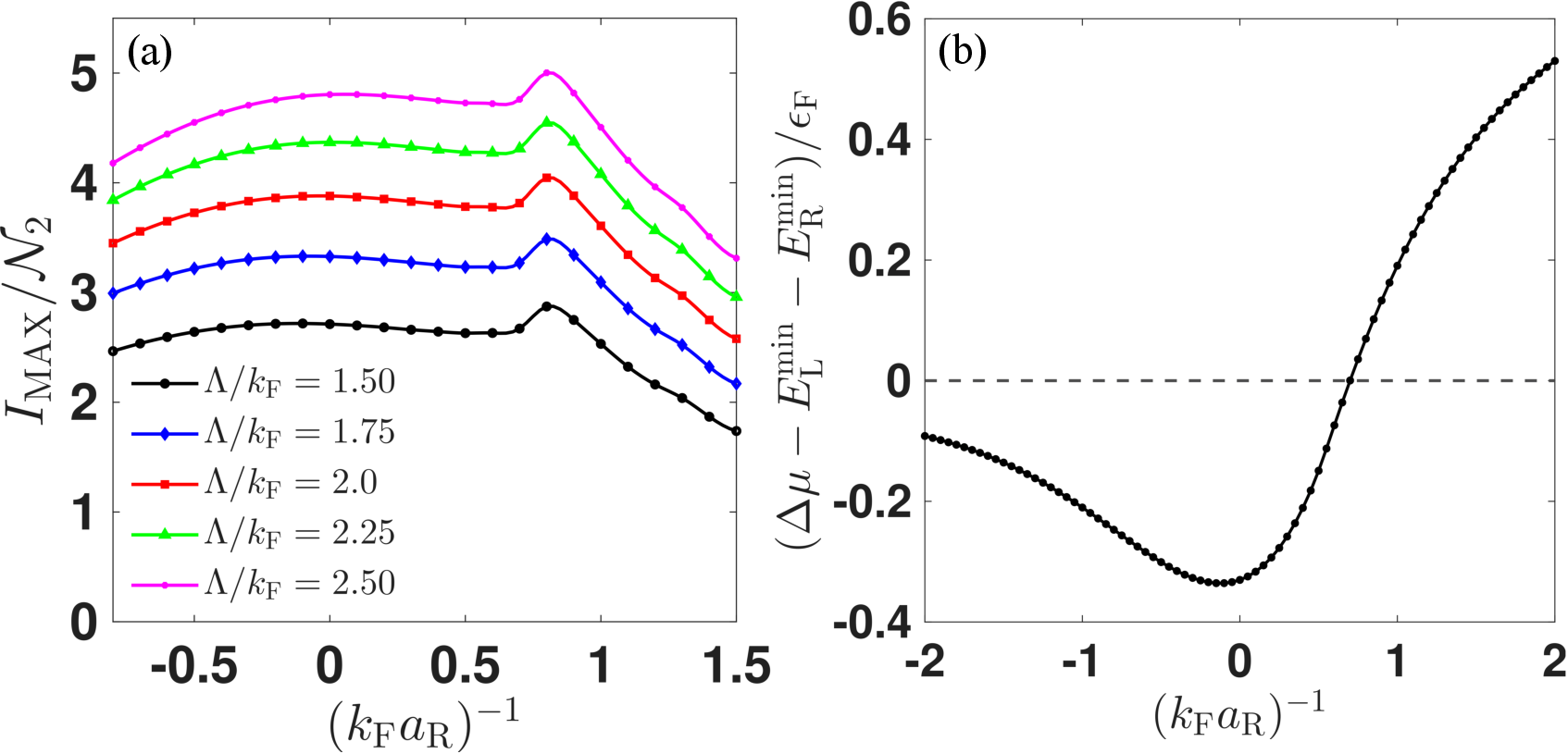}
    \caption{(a) The amplitude of AC Josephson current with different momentum cutoff $\Lambda$ when the right side changes from BCS to BEC regime, while the interaction strength in the left side is fixed as $(k_{\rm F}a_{\rm L})^{-1}=-1$. The left reservoir is fixed in BCS limit and the right side is tuned from BCS to BEC limit. A peak is found near $(k_{\rm F}a_{\rm R})^{-1}\approx0.8$, corresponding to the Riedel peak. (b) The value of $(\Delta\mu-E^{\rm min}_{\rm L}-E^{\rm min}_{\rm R})/\epsilon_{\rm F}$ as a function of $(k_{\rm F}a_{\rm R})^{-1}$. The zero point of this quantity supports that the Riedel-like enhancement in panel (a) appears when $\Delta \mu \approx E_{L}^{\rm min} + E_{R}^{\rm min}$.
    }
    \label{IMAX-V}
\end{figure}

To see how the magnitude of current, namely, the tunneling strength varies, we plot the current magnitude-interaction feature as shown in Fig.~\ref{IMAX-V} (a). While we used the same tunneling form factor as Eq.~(\ref{eq:20}), the bare tunneling coupling strength is treated as a property of the barrier. Therefore, our use of Eq.~(\ref{eq:20}) does not rely on expanding the tunneling amplitude with respect to a small chemical-potential difference. This assumption is appropriate as long as the barrier transmission is weak and does not vary strongly over the relevant energy window. A peak is found when the interaction strength in the right side reaches $(k_{\rm F}a_{\rm R})^{-1}\approx0.7$-$0.8$, which can be understood as a Riedel-like enhancement of the AC Josephson current~\cite{PhysRevLett.28.150,vernet1976direct,barone1982physics,PhysRevB.110.024517}. This Riedel peak appears when the effective chemical potential matches the threshold for quasiparticle excitations across the junction, namely
$\Delta\mu\approx E^{\rm min}_{\rm L}+E^{\rm min}_{\rm R}$. In our setup, the left reservoir is fixed in BCS limit where the chemical potential is positive. As the right side is tuned from BCS to BEC, its chemical potential gradually decreases and becomes negative, and this condition is equivalent to 
\begin{align}
    \Delta\mu\approx\left\{
    \begin{array}{cc}
        |\Delta_{\rm L}|+|\Delta_{\rm R}|  & \mu_{\rm R}>0  \\
        |\Delta_{\rm L}|+\sqrt{|\mu_{\rm R}|^2+|\Delta_{\rm R}|^2}  &  \mu_{\rm R}<0
    \end{array}
    \right. .
\end{align}
Fig.~\ref{IMAX-V} (b) shows the quantity of $\Delta\mu-(E^{\rm min}_{\rm L}+E^{\rm min}_{\rm R})$ as a function of $(k_{\rm F}a_{\rm R})^{-1}$, which crosses zero near $(k_{\rm F}a_{\rm R})^{-1}\simeq 0.7$-$0.8$, close to the peak position in Fig.~\ref{IMAX-V} (a). This supports the interpretation that the enhancement is a Riedel-like one associated with the resonance between $\Delta\mu$ and $E^{\rm min}_{\rm L}+E^{\rm min}_{\rm R}$.
This mechanism is directly analogous to the conventional Riedel singularity in superconducting tunnel junctions, where the AC Josephson current is resonantly enhanced near the threshold $eV=|\Delta_{\rm L}|+|\Delta_{\rm R}|$.

In the presence of a finite chemical potential bias induced by the interaction asymmetry,
one needs to distinguish the Josephson tunneling current $I_{\rm J}$ from the quasiparticle current $I_q$. However, $I_{\rm J}$ and $I_q$ can be unambiguously separated by their time dependencies. According to Eq.~\eqref{Iq}, the later is given by
\begin{align}
    I_q=8&\sum_{\bm{k},\bm{k}',\sigma}\mathcal{T}^2_{\bm{k},\bm{k}',\sigma}\int_{-\infty}^{\infty}\frac{d\omega}{2\pi}\operatorname{Im}G_{\bm{k},\sigma,{\rm L}}(\omega-\Delta\mu)\nonumber\\
    &\times\operatorname{Im}G_{\bm{k}',\sigma,{\rm R}}(\omega)[f(\omega-\Delta\mu)-f(\omega)],
\end{align}
which is time independent. Therefore, $I_q$ contributes only to the stationary background, whereas $I_{\rm J}$ is identified as the oscillatory part of the total current. Moreover, the different noise characteristics also help distinguish $I_{\rm J}$ from $I_q$. In an ideal ballistic limit, the Josephson current is associated with the coherent transfer of Cooper pairs without scattering, yielding an almost zero Fano factor (noise-current ratio)~\cite{PhysRevResearch.4.L042049}. On the other hand, $I_q$ originates from incoherent quasiparticle transport and generally carries a Fano factor $F\sim 1$~\cite{PhysRevB.96.094508}. This offers an additional natural criterion for separating the two contributions.

We finally estimate the absolute frequency scale of the AC Josephson oscillation. Restoring $\hbar$, the angular frequency is given by $\omega_{\rm J}=2|\Delta\mu|/\hbar$, or equivalently $f_{\rm J}=2|\Delta\mu|/h$. For a Fermi $^{6}\mathrm{Li}$ gas with typical density $n=10^{12}\,\mathrm{cm}^{-3}$, one obtains $k_{\rm F}\simeq3.1\times10^6\,\mathrm{m}^{-1}$ and $\epsilon_{\rm F}/h\simeq8.1\,\mathrm{kHz}$. In our mean-field calculation, when the left reservoir is fixed at $(k_{\rm F}a_{\rm L})^{-1}=-1$ and the right reservoir is tuned near the Riedel-like enhancement, $(k_{\rm F}a_{\rm R})^{-1}\simeq0.8$, the chemical-potential difference is of order $|\Delta\mu|\simeq1.3\,\epsilon_{\rm F}$. This gives
\begin{align}
    f_{\rm J} \simeq 2\times1.3\,\frac{\epsilon_{\rm F}}{h}
\simeq 2.1\times10^4\,\mathrm{Hz},
\end{align}
corresponding to a period $\tau_{\rm J}\simeq0.05\,\mathrm{ms}$. Therefore, for atomic lifetimes of a few milliseconds under interaction-tuning beams~\cite{PhysRevLett.122.040405}, several tens of AC Josephson oscillation periods could in principle occur near the Riedel-like enhancement. We have also noted that this estimate depends on density through $\epsilon_{\rm F}\propto n^{2/3}$, and that further improvements in lifetime and detection bandwidth would be beneficial for resolving the oscillatory current experimentally.

Although the present results are obtained within the BCS-BEC crossover mean-field theory, it would also be interesting to analyze the interaction-asymmetric Josephson junction by including beyond-mean-field corrections, for example within the extended $T$-matrix approximation (ETMA)~\cite{PhysRevA.95.043625}. A fully consistent beyond-mean-field treatment of the Josephson current would require not only replacing the mean-field values of $\Delta$ and $\mu$, but also using the corresponding dressed normal and anomalous Green's functions, as well as the associated spectral functions, in the tunneling-current formula. Such an extension would be left for future studies.

\section{Summary}\label{Sec5}

In this paper, based on nonequilibrium quantum field theory, we theoretical investigate the Josephson transport in a strongly-interacting Fermi gas with the two-terminal setup.
 We calculate the DC Josephson current through the interaction-symmetric junction where the interactions of both sides are tuned synchronously. To study the competition between increasing pair coherence and decreasing chemical potential, we evaluate the tunneling current in the entire the BCS-BEC crossover regime. A maximum of the DC Josephson current is found near the unitary limit as a result of the competition between the enhanced pairing gap and the reduced chemical potential. 
 Our numerical result is supported by the analytical expression of the DC Josephson current in the BCS and BEC limits.

We also analyze the Josephson current through an interaction-asymmetric junction, where one side is fixed in the BCS limit ($(k_{\rm F}a_{\rm R})^{-1}=-1$) and the other is tuned from the BCS to BEC limit. A Riedel peak is found when the chemical bias resonates with the sum of the minimum quasiparticle excitation energies on the two sides, indicating  that the Riedel-like singularity is not restricted to weak-coupling solid-state superconductors, but can also emerge in strongly interacting quantum gases with tunable interaction asymmetry. Our results can not only deepen the understanding of Josephson effect in strongly interacting regimes, but also be relevant to future experiments on engineered
quantum transport in ultracold atomic gases.

\begin{acknowledgements}

T.~Z. acknowledges support from HK GRF (Grant No.~17306024), CRF (Grants No.~C6009-20G, No.~C7012-21G, No.~C4050-23GF), CRS-HKU701/24 and a RGC Fellowship Award No.~HKU~RFS2223-7S03.
H.~T. was supported by the JSPS KAKENHI with Grants Nos.~JP22K13981 and JP23K22429.

\end{acknowledgements}

\appendix

\section{Asymptotic behaviors of DC Josephson current}\label{appendixA}

In the deep BEC limit, $\mu$ is negative and $|\Delta|\ll|\mu|$, The leading term of the DC current can be given by
\begin{align}
    I^{\rm BEC}_{\rm DC}\simeq\frac{4m^3|\Delta|^3\mathcal{T}^2\Lambda^4\sin(\Delta\phi)}{\pi^4}\mathcal{I}(\Lambda,\kappa),
\end{align}
where
\begin{widetext}
    \begin{align}
    \mathcal{I}(\Lambda,\kappa)\equiv&\int_0^\infty dk\int_0^\infty dk'\frac{k^2k'^2}{(k^2+\Lambda^2)(k'^2+\Lambda^2)(k^2+\kappa^2)(k'^2+\kappa^2)(k^2+k'^2+2\kappa^2)}\nonumber\\
    =&\frac{1}{(\Lambda^2-\kappa^2)^2}(\Lambda^4\mathcal{J}_{\Lambda\Lambda}-2\Lambda^2\kappa^2\mathcal{J}_{\Lambda\kappa}+\kappa^4\mathcal{J}_{\kappa\kappa})
\end{align}
with
\begin{align}\label{Jab}
    \mathcal{J}_{ab}=&\int_0^\infty dk\int_0^\infty dk'\frac{1}{(k^2+a^2)(k'^2+b^2)(k^2+k'^2+2\kappa^2)}\nonumber\\
    =&\frac{\pi}{2a}\int_0^\infty\frac{dk'}{(k'^2+b^2)\sqrt{k'^2+2\kappa^2}(a+\sqrt{k'^2+2\kappa^2})}
\end{align}
\end{widetext}
Setting $k'=\sqrt{2}\kappa\sinh t$, $\sqrt{k'^2+2\kappa^2}=\sqrt{2}\kappa\cosh t$, $dk'=\sqrt{2}\kappa\cosh t\,dt$, we have
\begin{align}
    \mathcal{J}_{ab}=\frac{\pi}{2a}\int_0^\infty\frac{dt}{(b^2+2\kappa^2\sinh^2 t)(a+\sqrt{2}\kappa\cosh t)}.
\end{align}
We further replace $t$ by $u=\tanh(t/2)$. Notice that $\sinh t=2u/(1-u^2)$, $\cosh t=(1+u^2)/(1-u^2)$ and $dt=2\,du/(1-u^2)$, and $\mathcal{J}_{ab}$ can be rewritten as
\begin{align}
    \mathcal{J}_{ab}=\frac{\pi}{a}&\int_0^1 du\frac{1-u^2}{b^2(1-u^2)^2+8\kappa^2u^2}\nonumber\\
    \times&\frac{1}{(a+\sqrt{2}\kappa)+(\sqrt{2}\kappa-a)u^2}.
\end{align}
We note that
\begin{align}
    &\frac{1}{b^2}\big[b^2(1-u^2)^2+8\kappa^2u^2\big]\nonumber\\
    =&\bigg[u^2+\bigg(\frac{\sqrt{2}\kappa+\sqrt{2\kappa^2-b^2}}{b}\bigg)^2\bigg]\nonumber\\
    &\times\bigg[u^2+\bigg(\frac{\sqrt{2}\kappa-\sqrt{2\kappa^2-b^2}}{b}\bigg)^2\bigg]\nonumber\\
    =&(u^2+\rho_+^2)(u^2+\rho_-^2).
\end{align}
Therefore the integral Eq.~\ref{Jab} is rewritten as 
\begin{align}
    \mathcal{J}_{ab}=\frac{\pi}{ab^2}&\int_0^1 du\frac{1-u^2}{(u^2+\rho_+^2)(u^2+\rho_-^2)}\nonumber\\
    \times&\frac{1}{(a+\sqrt{2}\kappa)+(\sqrt{2}\kappa-a)u^2},
\end{align}
which can be decomposed as 
\begin{align}
    &\frac{1-u^2}{(u^2+\rho_+^2)(u^2+\rho_-^2)[(a+\sqrt{2}\kappa)+(\sqrt{2}\kappa-a)u^2]}\nonumber\\
    =&\frac{C_+}{u^2+\rho_+^2}+\frac{C_-}{u^2+\rho_-^2}+\frac{C_0}{a+\sqrt{2}\kappa+(\sqrt{2}\kappa-a)u^2}.
\end{align}
Thus the integral becomes
\begin{align}
    \mathcal{J}_{ab}=\frac{\pi}{ab^2}\bigg[&C_+\int_0^1\frac{du}{u^2+\rho^2_+}+C_-\int_0^1\frac{du}{u^2+\rho^2_-}\nonumber\\
    &+C_0\int_0^1\frac{du}{a+\sqrt{2}\kappa+(\sqrt{2}\kappa-a)u^2}\bigg].
\end{align}
We focus on the deep BEC limit, where we typically have $\kappa>a$ and automatically $\sqrt{2}\kappa>a$. Then the simplified form is given by
\begin{align}
    \mathcal{J}_{ab}=\frac{\pi}{ab^2}\Big[C_+&\frac{1}{\rho_+}\arctan\frac{1}{\rho_+}+C_-\frac{1}{\rho_-}\arctan\frac{1}{\rho_-}\nonumber\\
    &+C_0\Phi(a,\kappa)\Big],
\end{align}
where 
\begin{align}
    \Phi(a,\kappa)=\frac{\arctan\sqrt{\frac{\sqrt{2}\kappa-a}{\sqrt{2}\kappa+a}}}{\sqrt{(\sqrt{2}\kappa-a)(\sqrt{2}\kappa+a)}}.
\end{align}
Inserting $\{a,b\}=\{\Lambda,\kappa\}$ into the closed form, we obtain 
\begin{align}
    \mathcal{J}_{\kappa\kappa}=\frac{\pi(\pi-2)}{8\kappa^4},
\end{align}
\begin{align}
    \mathcal{J}_{\Lambda\kappa}=\frac{\pi^2(\Lambda-2\kappa)}{8\Lambda\kappa^2(\Lambda^2-\kappa^2)}+\frac{\pi\arctan{\sqrt{\frac{\sqrt{2}\kappa-\Lambda}{\sqrt{2}\kappa+\Lambda}}}}{\Lambda(\Lambda^2-\kappa^2)\sqrt{2\kappa^2-\Lambda^2}},
\end{align}
\begin{align}
    \mathcal{J}_{\Lambda\Lambda}=\frac{\pi(8\Lambda\theta-2\pi\Lambda+\pi\sqrt{2\kappa^2-\Lambda^2})}{8\Lambda^2\sqrt{2\kappa^2-\Lambda^2}(\kappa^2-\Lambda^2)},
\end{align}
where $\theta=\arctan\big(\frac{\Lambda}{\sqrt{2}\kappa+\sqrt{2\kappa^2-\Lambda^2}}\big)$.

In the BCS limit the DC Josephson current can be approximately given by Eq.~(\ref{IBCS}).We take $\xi=|\Delta|\sinh{u}$ and $\xi'=|\Delta|\sinh{v}$. Then $E=|\Delta|\cosh{u}$ and $E'=|\Delta|\cosh{v}$. We have $d\xi=|\Delta|\cosh{u}\,du$, and Eq.~(\ref{IBCS}) becomes
\begin{align}
    I_{\rm DC}^{\rm BCS}\simeq 4\mathcal{T}_{\rm F}^2\sin(\Delta\phi)\int_{-\infty}^{\infty} du \int_{-\infty}^{\infty} dv\frac{N(0)^2|\Delta|}{\cosh{u}+\cosh{v}}.
\end{align}
Now taking $u=p+q$ and $v=p-q$, we have
\begin{align}
    \cosh{u}+\cosh{v}=2\cosh{p}\cosh{q}\cr
    du\,dv=\bigg|\frac{\partial(u,v)}{\partial(p,q)}\bigg|dp\, dq=2dp\,dq,
\end{align}
and the Josephson current in the BCS limit becomes
\begin{align}
    I_{\rm DC}^{\rm BCS}\simeq& 4\mathcal{T}_{\rm F}^2\sin(\Delta\phi)\int_{-\infty}^{\infty} dp \int_{-\infty}^{\infty} dq\frac{N(0)^2|\Delta|}{\cosh{p}\cosh{q}}\cr
    =&4\pi^2 N(0)^2\mathcal{T}_{\rm F}^2|\Delta|\sin(\Delta\phi).
\end{align}

\bibliography{ref.bib}

\end{document}